\documentclass[preprint]{elsarticle}

\usepackage{lineno,hyperref}
\modulolinenumbers[5]

\journal{SBMF 2019}

\usepackage{amsmath,amssymb,amsfonts}
\usepackage{cspsymb}
\usepackage{stfloats}
\usepackage{algorithmic}
\usepackage{graphicx}
\usepackage{textcomp}
\usepackage{xcolor}
\usepackage{rotating}
\usepackage{url}

\begin{document}

\begin{frontmatter}

\title{A framework for verifying deadlock and nondeterminism in UML activity diagrams based on CSP}

\author{Lucas Lima, Amaury Tavares and Sidney C. Nogueira}
\address{Departamento de Computa\c{c}\~ao - Universidade Federal Rural de Pernambuco \\ Rua Dom Manuel de Medeiros, s/n, Dois Irm\~aos - CEP: 52171-900 - Recife/PE}


\cortext[mycorrespondingauthor]{Corresponding authors}
\ead{lucas.albertins@ufrpe.br, amaury.tavares@ufrpe.br, sidney.nogueira@ufrpe.br}


\begin{abstract}
Deadlock and nondeterminism may become increasingly hard to detect in concurrent and distributed systems. UML activity diagrams are flowcharts that model sequential and concurrent behavior. Although the UML community widely adopts such diagrams, there is no standard approach to verify the presence of deadlock and nondeterministic behavior in activity diagrams. Nondeterminism is usually neglected in the literature even though it may be considered a very relevant property. This work proposes a framework for the automatic verification of deadlock and nondeterminism in UML activity diagrams. It introduces a compositional CSP semantics for activity diagrams that is used to automatically generate CSP specifications from UML models. These specifications are the input for the automatic verification of deadlock and nondeterministic behavior using the FDR refinement checker. We propose a plugin for the Astah modeling environment that mechanizes the translation process, and that calls FDR in the background to perform the verification of properties. The tool keeps the traceability between a diagram and its CSP specification. It parses the FDR results to highlight the diagram paths that lead to a deadlock or a nondeterministic behavior. This framework adds verification capabilities to the UML modeling tool and keeps the formal semantics transparent to the users. Therefore, the user does not need to understand or manipulate formal notations during modeling. We present the results of a case study that applies the proposed framework for the verification of models in the domain of cloud computing. We discuss future applications due to the potential of our approach.
\end{abstract}


\end{frontmatter}


\section{Introduction}
\label{sec:introduction}


The search for solutions to improve the detection of failures is still a relevant topic because systems have become considerably intricate. Several works portray the concern for early failure detection due to the high cost of fixing bugs in later stages of development~\cite{Boehm1981,Haskins04,shull2002}. Moreover, the increasing complexity of systems has escalated the difficulty in assuring acceptable quality standards. If we consider critical
systems, for which safety is a major concern, early Verification and Validation (V\&V) is recognized as a valuable approach to promote dependability. Therefore, in addition to the standard effort in testing tasks as the system is being constructed, several other approaches are being proposed to verify the design of systems in order to identify possible flaws.

In the last decades we have seen a considerable growth of a category of systems that are available online. Cloud computing has emerged as a model for deploying systems and services on the Internet or private networks instead of local environments. It has been embraced by major IT companies such as Amazon, Apple, Google, HP, IBM, Microsoft, Oracle, and others. Cloud
service providers have their own service infrastructures and are responsible for infrastructure and service management. However, this management of resources can become very complex due to the challenge to coordinate concurrent computations. Often the parallel computation involves multiple stages, and all concurrent activities must finish one stage before starting the execution of the next one.
One potential problem for concurrent execution of multiple tasks is the presence of deadlocks~\cite{Marinescu2013}. A concurrent system is deadlocked if no task can make any progress. This usually happens because each task is waiting for communication with others~\cite{Roscoe1997}. Another aspect of concern is the possibility of nondeterministic choices. Nondeterminism is important for abstraction purposes, for underspecification, and to model the interface with an unknown or unpredictable environment~\cite{Baier2008}.

In terms of system and software development, UML (Unified Modeling Language)~\cite{UML} is considered a standard language for modeling and designing software systems. There is a wide availability of tooling support for dealing with its models providing capabilities not only for the modeling perspective but also features like code generation, communication and verification of the designs~\cite{ibm,PTC,Microsoft,Astah}. UML provides several diagrams to express structural and behavioral aspects of systems, however, a small subset of these diagrams are used in large scale. According to~\cite{Reggio13}, the activity diagram is one the most used among all UML diagrams. It is a behavioral diagram focused on the description of a coordinated dynamics with emphasis on the
sequence of actions and conditions. The flows of an activity diagram can be used just for controlling the order that actions are executed or for representing the communication of data as well. Usually they are used for both high-level designs, e.g. modeling business process, and low-level designs, for instance, to describe an algorithm to be implemented. 

As one of the main UML diagrams used in practice, several works propose methods for validating it in order to improve early error detection. These works~\cite{Machida11,Baldan05,Banti2011,elmansouri_uml_2011,Eshuis2006,George2012,Alawneh2006,Abdelhalim2010,Ouchani2014} provide relevant contributions to the assessment of activity diagrams, however, it is not clear the level of automation and integration with modeling environments. Despite some of the works approach the verification of deadlock freedom, nondeterminism is usually neglected. It is, however,
especially important in notations for refinement, where nondeterminism is used for abstraction. Normally, they translate the activity diagrams to some formal notation (for instance, NuSMV~\cite{NuSMV} or PRISM~\cite{PRISM}) and then perform some kind of assessment using formal methods.

While UML is well-suited for designing systems in general, it lacks support for reasoning. On the other hand, formal models require understanding and manipulation of mathematical concepts, however, they are unambiguous and amenable to reasoning. Therefore, we plan to bring the best of both methodologies together. Most of the initiatives ~\cite{Eshuis2006,elmansouri_uml_2011,Alawneh2006,Baldan05,Abdelhalim2010,Ouchani2014} use model checking to verify properties. Model checking is an automatic verification technique that explores all possible system states in order to check the satisfiability of a given property~\cite{Baier2008}. 

We use the process algebra CSP (Communicating Sequential Processes)~\cite{Hoa85} as the underlying semantic domain for activity diagrams. CSP can be used to describe systems composed of interacting elements, which are independent self-contained processes with interfaces used to interact with the environment. CSP specifications can be analyzed by the FDR tool~\cite{FDR3}, which is a well-known model checker for CSP. Basically, FDR translates a CSP specification to a Labelled Transition System (LTS) and traverses the states of this model in order to check a specific property. We choose the CSP semantics for activity diagrams for several reasons. First, the expressiveness of the CSP operators allows us to define a compositional semantics where each constructor is specified independently of the others. This approach facilitates the implementation and maintainability of our semantics. Moreover, FDR is a mature CSP model checker that provides deadlock and determinism checking. Finally, we can use FDR for checking refinement, which we plan to explore in the future.

Our work aims to validate absence of deadlock and nondeterminism in activity diagrams using an automated approach. Although we use formal methods to validate the models, this work has, by principle, to provide a framework that does not require any knowledge on formal techniques or rigorous notations. This is accomplished by creating a shell on a modeling tool that hides all formal aspects from the user. The Astah tool~\cite{Astah} is a very popular modeling environment that supports the authoring of activity diagrams as well as other UML diagrams. Such a tool has got extension capabilities that facilitate the creation of plug-ins and ease the integration with other tools. Our framework is built as a plug-in for the Astah modeling environment. After creating their activity diagrams, users can select options for verifying deadlock freedom or checking determinism. Our plug-in translates the activity diagram model to CSP and performs the selected assertion on the translated model via integration with FDR. In case a deadlock or a nondeterministic choice is found, FDR delivers a counterexample, which is translated back to an activity diagram highlighted with the undesirable path. Thus, all formal notation and reasoning process is hidden from the user. In summary, our main contributions are:

\begin{itemize}
    \item A suitable process algebraic semantics for UML activity diagrams covering a considerable amount of constructors.
    \item An automatic method to translate activity diagram models to CSP specifications.
    \item An integration with FDR to automatically assert the absence of deadlock and nondeterminism.
    \item A traceability mechanism to provide FDR counterexamples in terms of activity diagrams.
\end{itemize}

\begin{figure}[!htp]
\centering
\includegraphics[scale=0.35]{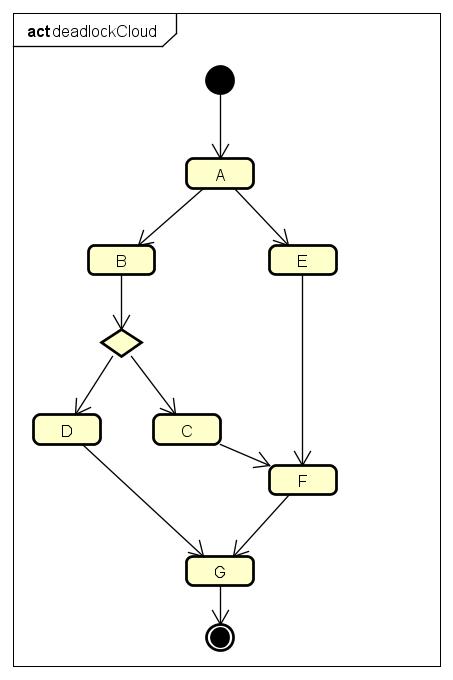}
\caption{An activity diagram with a deadlock~\cite{Marinescu2013}.}
\label{fig:deadlockedad}
\end{figure}

Figure~\ref{fig:deadlockedad} illustrates an example adapted from~\cite{Marinescu2013} in the context of cloud computing. In this example, several tasks are executed with several dependencies among them. Some of them are executed concurrently as well. There is a deadlock in this scenario that could go unnoticed at first glance. If task D is chosen, then F will never be instantiated, because it requires the completion of both C and E. The process will never terminate, because G requires completion of both D and F. Our approach detects this possible deadlock and marks the path that leads to it.

The remainder of this paper is organized as follows. Section~\ref{sec:background} presents the basic concepts used by our work. Section~\ref{sec:semantics} introduces the activity diagram semantics in terms of CSP. Section~\ref{sec:toolsupport} describes the tooling support we provide. Section~\ref{sec:casestudy} illustrates our verification strategies in some case studies. Related works are presented in Section~\ref{sec:relatedwork} and Section~\ref{sec:conclusions} concludes and discusses initiatives for future work.  

\section{Background}
\label{sec:background}

We briefly introduce the key concepts about UML Activity diagrams and
the CSP process algebra.

\subsection{UML Activity Diagram}

\begin{figure}[h]
\centering
\includegraphics[scale=0.37]{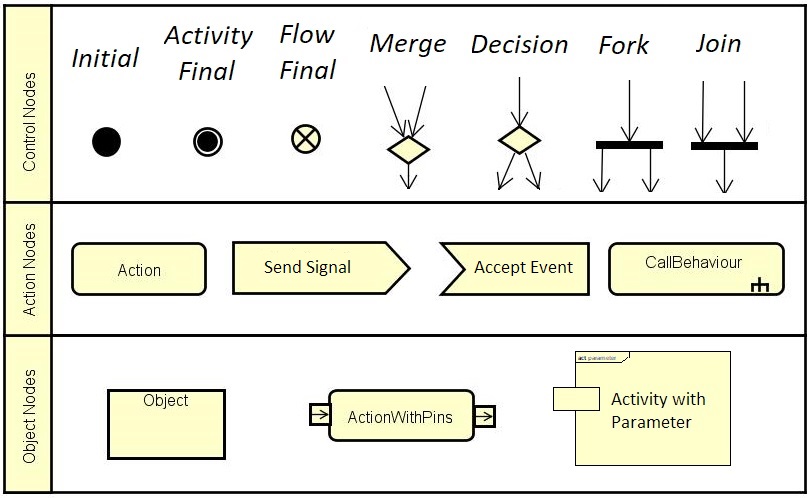}
\caption{Activity diagram nodes.}
\label{fig:nodes}
\end{figure}

A UML activity diagram is a graph of activity nodes interconnected by activity edges~\cite{UML}. An activity node can be either an action node, an object node or a control node. Activity edges are directed connections between two activity nodes, they can be either a control flow, which is used to explicitly sequence execution of activity nodes, or an object flow, which can have data (objects) passing along it. An action node executes a desired behavior when ready, including sending or receiving signals or invoking another activity. An object node explicitly holds objects that arrive in its incoming edges and offers them to the outgoing edges. Control nodes organize the sequencing of flows. They act as ``traffic switches" across the activity edges. Figure~\ref{fig:nodes} shows all types of control nodes (initial, activity final, flow final, merge, decision, fork and join), some types of action nodes (basic action, send signal, accept event and call behavior), and three types of object nodes (basic object, pins and parameter). The semantics for each constructor is described textually in the UML specification~\cite{UML}.

An example of an activity diagram is displayed in Figure~\ref{fig:deadlockedad} (Section~\ref{sec:introduction}). It has three control nodes (initial, decision and activity final) and seven basic action nodes. Besides the semantics of each node, the execution semantics of an activity diagram is described in terms of tokens flowing through the edges and nodes. Activity edges are directed with tokens flowing from the source activity node to the target activity node. However, the token must only flow if the target is ready to accept it. Some nodes may generate tokens, for instance, an initial node creates tokens on its outgoing edges when the activity starts. Other nodes only consume tokens, like flow final and activity final nodes. Object nodes can hold several tokens before passing them to the following nodes. An action node can only be executed once all incoming edges are offering tokens, and when it terminates, it must offer tokens in its outgoing edges.

Finally, an activity diagram can only terminate in two scenarios: if there are no more active tokens flowing through the activity after it has been started, or, if an activity final node has consumed a token. In the latter case, all current flows are halted. 

\subsection{CSP}
\label{sec:csp}

The CSP process algebra is very expressive to specify systems composed of
interacting components, which are independent self-contained processes
with interfaces used to interact with the environment. Such a formalism
provides constructs to explicitly specify and reason about interactions
between different components. 

In CSP, a process is the basic unit for describing behavior. It is defined in terms of events and other processes. The function $\alpha(P)$ yields the alphabet of a process $P$, that is, the events that the process $P$ may communicate. 
The primitive process $SKIP$ 
represents successful termination. The primitive process $STOP$ represents the canonical deadlock. A process $a \then P$ offers the event $a$ to the
environment and then behaves as the process $P$. CSP channels abstract a set of events with a common prefix. 
The syntax $a?x$ represents the channel $a$ inputs $x$, whose value is chosen by the environment. The syntax $a.x$ ($a!x$) represents a value $x$ communicated by the channel $a$. The difference between $.$ and $!$ is very subtle. For instance, consider $a$ is a channel of type $A.B.C.D$ then, the communication $a?x.y!z.t$ is equivalent to $a?x?y!z!t$, because a dot following a $?$ ($!$) is taken to be part of a pattern that is matched by the input.
The sequential composition $P1;P2$ behaves like the process $P1$
and, provided it terminates successfully, $P2$ takes over.
The CSP notation has no explicit operator for recursion, but it allows one to use the
name of the process in its definition. For instance, the process $P = a \then P$ communicates the event $a$ and then behaves as $P$.

The external choice $P1 \extchoice P2$ initially offers
events of both processes $P1$ and $P2$. The communication of the first event resolves the choice in favour of the process that performs it. The environment has no control over the internal (nondeterministic) choice $P1 \intchoice P2$: the process internally chooses to behave as $P1$ or $P2$.
The parallel composition $P1 \parallel[cs] P2$ synchronizes
$P1$ and $P2$ on the events in the set $cs$;
events not in $cs$ occur independently. Processes
composed in interleaving $P1 \interleave P2$ progress without synchronization. The event
hiding operator $P \hide cs$ internalizes the events that
belong to the set $cs$, which become no longer visible to
the environment. The interruption operator ($\interrupt$) allows a process to be interrupted by another. The process $P \interrupt Q$ behaves as $P$ until $Q$ communicates an event. When this happens, we say that $P$ has been interrupted by $Q$.

CSP has a very mature tool support. This is one of the reasons for the success of CSP in industry. 
For instance, the FDR tool~\cite{FDR3} verifies process refinement as well as properties like deadlock, divergence and nondeterminism. In this work, we focus on the detection of deadlock and nondeterminism. FDR takes as input specifications in CSP$_M$, a machine readable version of CSP and translates them to LTL (Labelled Transition System), a state machine based notation. The approach taken by FDR for checking properties like deadlock and nondeterminism is based on global analysis, where the
entire model is expanded and exhaustively checked. However, FDR has several optimization mechanisms in order to improve performance of its reasoning mechanism~\cite{Roscoe1997}. 

\section{A formal semantics for UML activity diagrams}
\label{sec:semantics}

This section introduces the formal semantics for UML activity diagrams using the CSP notation. This semantics is an evolution from previous works~\cite{Lima2013,Lima2015,LIMA2016}. 
Here, we do not provide a complete description of it, instead, we give an intuition on how the elements are translated to CSP. We emphasize that our semantics preserves compositionality of the activity diagram constructors, that is, each element has a CSP representation which is independent of other elements in the diagram. This is achieved by the compositionality of the CSP parallelism operators. This feature facilitates both mechanization of the translation and traceability from CSP traces to activity diagram elements.

\begin{figure*}[h]
\centering
\includegraphics[scale=0.42]{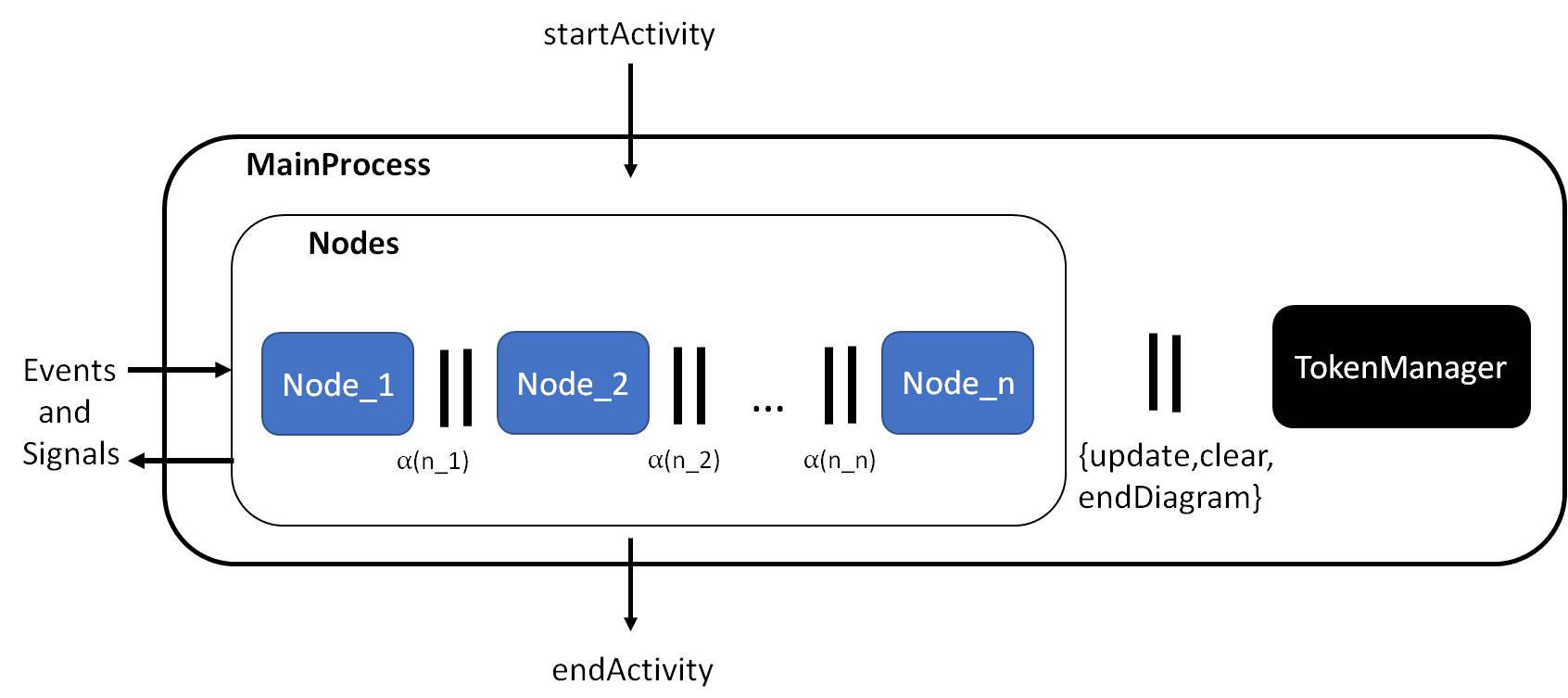}
\caption{Activity diagram semantics in CSP.}
\label{fig:activitycsp}
\end{figure*}

According to the UML specification~\cite{UML}, an activity is described in terms of nodes and edges between them. In our semantics, these elements are represented in CSP by processes and events, respectively. Events in CSP can be communicated by channels. We use two types of channels to represent edges: $ce$ for control edges events and $oe$ for object edges events. The activity is represented by a CSP process as well. Figure~\ref{fig:activitycsp} graphically illustrates the representation of an activity in CSP. The boxes represent CSP processes. An activity starts once the $startActivity$ event is communicated. This event may receive any input data needed in the activity parameter nodes. The $startActivity$ event is followed by a process that composes in parallel the processes for all activity nodes (box $Nodes$ in Figure~\ref{fig:activitycsp}). The process for each node synchronizes on the events related to its edges. For instance, if there is a control edge between two nodes, the event $ce.n$, where $n$ is the index of the edge, is part of the synchronization alphabet of the processes from both nodes. In order to keep the amount of active tokens to control the termination of the activity, the $Nodes$ process is composed in parallel with an auxiliary process called  $Token\_Manager$.
Once an activity terminates, the main process communicates the $endActivity$ event with any data in output parameter nodes. While an activity has active flows it may communicate events and signals with other activities, for instance, to invoke another activity using a call behavior action.

The underlying semantics of activity diagrams is described by the flow of tokens from a node to another. These tokens trigger the execution of nodes and also cause the termination of the diagram. An activity must terminate once there is no active token or a token reaches an activity final node. In order to comply with this constraint, the \textit{Token\_Manager} process controls the termination of the diagram by maintaining the number of active tokens. This process is shown next.

\vspace{1ex}


\noindent$Token\_Manager(n,init) =$ \\
\hspace*{1cm}$update?x \then Token\_Manager(n+x,true)$\\
\hspace*{0.7cm}$\extchoice$ $clear \then endDiagram \then SKIP$\\
\hspace*{0.7cm}$\extchoice$ $(n == 0$ $and$ $init)$ $\&$ $endDiagram \then SKIP$

\vspace{1ex}

The initial values of $n$ and $init$ are 0 (zero) and $false$, respectively. The first parameter stands for the number of active tokens, the second one states whether the diagram is active. This process can receive communications on the channel $update$ with an integer value $x$ to update the current number of tokens to $n+x$. The value of $x$ is positive when the number of active tokens increases (e.g. after a fork node is performed); the value is negative when the number of active tokens decreases (e.g. after a join node is performed). After the first $update$ event, the $Token\_Manager$ becomes active, so the value of $init$ is set to $true$. An activity final node terminates the activity once a token has reached it. In our semantics it is the only node that communicates the event $clear$, which synchronizes with the $Token\_Manager$ process. Once it has been synchronized, the flows of the diagram terminate (event $endDiagram$). Another possibility of termination is when the number of tokens reaches 0 (zero). When this happens, once the diagram has already started ($init$ is $true$), the flow of the diagram terminates as well. After the event $endDiagram$ is communicated, all CSP processes of the nodes are interrupted and the $endActivity$ event is performed.

As described in Section~\ref{sec:background}, the types of nodes are action, control and object. Tables~\ref{tab:basic-action-csp} --- \ref{tab:call-behavior-action-csp}~ show the mapping of nodes to CSP for action nodes and Tables~\ref{tab:initial-csp} --- \ref{tab:decision-csp}~describe the mapping for control nodes. For simplicity, these mappings illustrate nodes with control edges only, which are indexed by integers (shown as $m,n,u,t$ and $i$), and ellipsis denotes an interval of edges. In case of object edges are used instead, they would communicate data on the channel $oe$. We also omit the 
In the following, we overview the mappings. 

Table~\ref{tab:basic-action-csp} shows the CSP semantics for a basic action node. Let $ActionX$ be a CSP process that represents a basic action node. Initially, this process receives communications on all incoming edge events ($ce.m \then SKIP \interleave ... \interleave ce.n \then SKIP$), which are in interleaving. This is followed by the execution of its behavior. The behavior of a basic action (represented in Table~\ref{tab:basic-action-csp} as the process $BehaviorActionX$) can be to communicate an event labelled as the action name, and, optionally, it may manipulate data received on its input pins and assignments to its output pins. Afterwards, an action node communicates in interleaving all events related to its outgoing edges ($ce.u \then SKIP \interleave ... \interleave ce.t \then SKIP$). 

We omitted in Table~\ref{tab:basic-action-csp} how the activity node is interrupted when the diagram is concluded. Consider the process $ActionX$ shown in Table~\ref{tab:basic-action-csp}. There is a process $ActionX\_t = ActionX \interrupt (endDiagram \then SKIP)$ that is interrupted by an event $endDiagram$, which is controlled by the $Token\_Manager$ process. This mechanism allows the nodes to be executed as long as there are active flows in the diagram. In fact, the process $ActionX\_t$ is the one that is composed in parallel with the processes that represents other nodes. Every kind of node presented in this section has a process following the same pattern to allow the interruption of its flow. For conciseness, we omit this second process for all nodes discussed in this section.
\begin{table}[!ht]
\caption{Basic Action --- CSP Semantics}
\centering
\begin{tabular}{|c|}
	\hline
	\begin{tabular}{c}
	\includegraphics[scale=0.45]{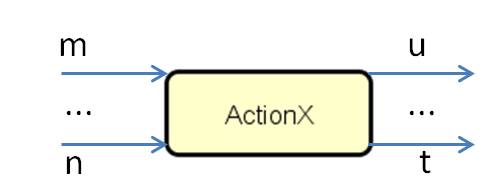}  
	\end{tabular} 
    \\
    \hline
    \begin{tabular}{ll}
        $ActionX = $ & $(ce.m \then SKIP \interleave ... \interleave ce.n \then SKIP);$\\
        			& $ BehaviorActionX;$  \\  
        			& $(ce.u \then SKIP \interleave ... \interleave ce.t \then SKIP);$\\
        			& $ActionX$ 
    \end{tabular}\\
    \hline
\end{tabular}
\label{tab:basic-action-csp}
\end{table}

A sending signal action (Table~\ref{tab:send-signal-action-csp}) communicates the signal event (described in its action) with values for the sender and receiver (event $signalX!sender?target$ in Table~\ref{tab:send-signal-action-csp}). The receiver can be any activity waiting for the signal. In a similar way, an accept event action (Table~\ref{tab:accept-event-action-csp}) communicates a signal event, however, the sender depends on which activity synchronizes the accepted event. For instance, given that two activities are running in parallel, at certain point one can send a signal through a send signal action and another receives this signal by the accept event action. Both activities synchronize on this specific event. 
\begin{table}[!ht]
\caption{Send Signal Action --- CSP Semantics}
\centering
\begin{tabular}{|c|}
	\hline
	\begin{tabular}{c}
	\includegraphics[scale=0.45]{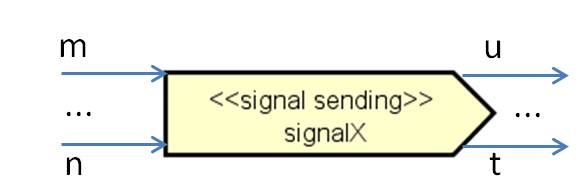}  
	\end{tabular} 
    \\
    \hline
    \begin{tabular}{ll}
        $SendSignalActionX = $ & $ (ce.m \then SKIP \interleave ... \interleave ce.n \then SKIP);$  \\
        & $signalX!sender?target \then SKIP;$ \\      
        & $(ce.u \then SKIP \interleave ... \interleave ce.t \then SKIP);$\\
        & $SendSignalActionX$ 
    \end{tabular}\\
    \hline
\end{tabular}
\label{tab:send-signal-action-csp}
\end{table}
\begin{table}[!ht]
\caption{Accept Event Action --- CSP Semantics}
\centering
\begin{tabular}{|c|}
	\hline
	\begin{tabular}{c}
	\includegraphics[scale=0.45]{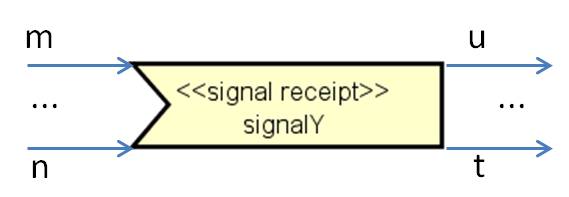}  
	\end{tabular} 
    \\
    \hline
    \begin{tabular}{ll}
        $AcceptEventActionY = $ & $(ce.m \then SKIP \interleave ... \interleave ce.n \then SKIP);$  \\
        & $signalY?sender!target \then SKIP;$ \\      
        & $(ce.u \then SKIP \interleave ... \interleave ce.t \then SKIP);$\\
        & $AcceptEventActionY$ 
    \end{tabular}\\
    \hline
\end{tabular}
\label{tab:accept-event-action-csp}
\end{table}

A call behavior action (Table~\ref{tab:call-behavior-action-csp}) can invoke the behavior of another activity by synchronizing on its $startActivity$ event and waiting for its termination until it synchronizes on its $endActivity$ event. Each activity has a particular event to represent the start and the end of the activity. Thus, these events are suffixed by the activity identification (e.g.  $startActivityW$ and $endActivityW$).
\begin{table}[!ht]
\caption{Call Behavior Action --- CSP Semantics}
\centering
\begin{tabular}{|c|}
	\hline
	\begin{tabular}{c}
	\includegraphics[scale=0.45]{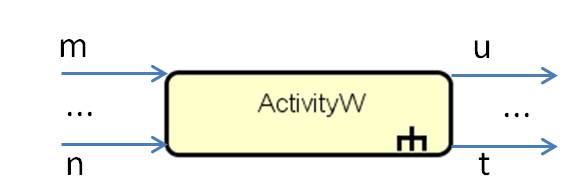}  
	\end{tabular} 
    \\
    \hline
    \begin{tabular}{ll}
        $CallBehaviorActionW = $ & $ (ce.m \then SKIP \interleave ... \interleave ce.n \then SKIP);$  \\
        & $startActivityW \then endActivityW \then SKIP;$ \\      
        & $(ce.u \then SKIP \interleave ... \interleave ce.t \then SKIP);$\\
        & $CallBehaviorActionW$ 
    \end{tabular}\\
    \hline
\end{tabular}
\label{tab:call-behavior-action-csp}
\end{table}

Control nodes are used to organize the control and object flows of the activity. One of their particular features is the generation/consumption of tokens. Initial and fork nodes (Tables \ref{tab:initial-csp} and \ref{tab:fork-csp}) increase the number of active tokens by synchronizing on the channel $update$ with the $Token\_Manager$ process communicating the number of flows to be increased according to the number of outgoing edges ($\#outEdges$). Flow final and join nodes (Tables \ref{tab:flow-final-csp} and \ref{tab:join-csp}) decrease the number of tokens using the same channel. However, flow final nodes decrement one token at a time while this number for join nodes is calculated according to the number of incoming edges ($\#incEdges$). Activity final nodes (Table \ref{tab:activity-final-csp}) terminate all flows after the first communication on any of its incoming edges because it synchronizes on the $clear$ event with the $Token\_Manager$ process. Merge nodes (Table \ref{tab:merge-csp}) wait communication on any of its incoming edges using the external choice operator, and then, communicates the event related to its outgoing edge. Decision nodes (Table \ref{tab:decision-csp}) wait communication of its incoming edge event, then it evaluates the guards of its outgoing edges in order to decide which path should be chosen. When more than one guard is true, then the choice on the outgoing edge is non-deterministic. If no guards are true than a deadlock happens. 

Some simplifications are made in the presentation of action and control nodes
to improve readability. For instance, decision nodes may have an additional incoming edge stereotyped $decisionInputFlow$ that provides data to be evaluated by the guards and it is considered in the formal semantics.
\begin{table}[!ht]
\caption{Initial --- CSP Semantics}
\centering
\begin{tabular}{|c|}
	\hline
	\begin{tabular}{c}
	\includegraphics[scale=0.45]{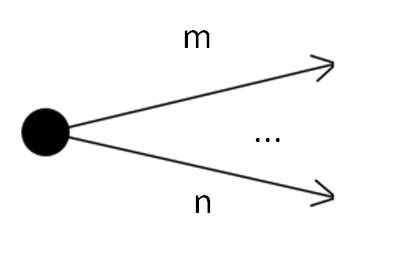}  
	\end{tabular} 
    \\
    \hline
    \begin{tabular}{c}
        $Init = update.(\#outEdges) \then (ce.m \then SKIP \interleave ... \interleave ce.n \then SKIP)$  \\ 
    \end{tabular}\\
    \hline
\end{tabular}
\label{tab:initial-csp}
\end{table}
\begin{table}[!ht]
\caption{Fork --- CSP Semantics}
\centering
\begin{tabular}{|c|}
	\hline
	\begin{tabular}{c}
	\includegraphics[scale=0.45]{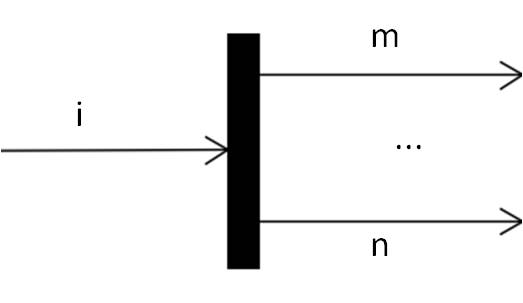}  
	\end{tabular} 
    \\
    \hline
    \begin{tabular}{ll}
        $Fork = $ & $ce.i \then update.(\#outEgdes-1) \then $\\
                  & $(ce.m \then SKIP \interleave ... \interleave ce.n \then SKIP); Fork$  \\
    \end{tabular}\\
    \hline
\end{tabular}
\label{tab:fork-csp}
\end{table}

\begin{table}[!ht]
\caption{Flow Final --- CSP Semantics}
\centering
\begin{tabular}{|c|}
	\hline
	\begin{tabular}{c}
	\includegraphics[scale=0.45]{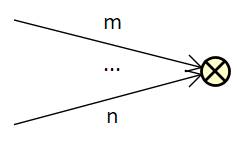}  
	\end{tabular} 
    \\
    \hline
    \begin{tabular}{ll}
        $FlowFinal = $ & $(ce.m \then SKIP \extchoice ... \extchoice ce.n \then SKIP);$\\
                       & $update.(-1) \then FlowFinal$  \\
    \end{tabular}\\
    \hline
\end{tabular}
\label{tab:flow-final-csp}
\end{table}
\begin{table}[!ht]
\caption{Join --- CSP Semantics}
\centering
\begin{tabular}{|c|}
	\hline
	\begin{tabular}{c}
	\includegraphics[scale=0.45]{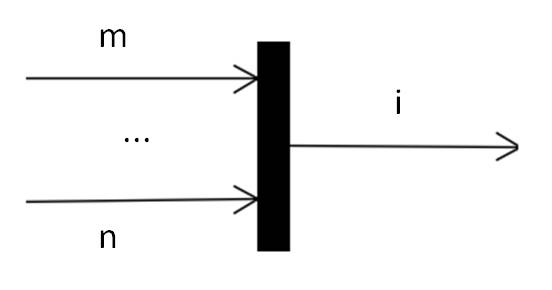}  
	\end{tabular} 
    \\
    \hline
    \begin{tabular}{ll}
        $Join = $ & $ (ce.m \then SKIP \interleave ... \interleave ce.n \then SKIP);$  \\
        		& $(update.(1-\#incEdges) \then ce.i \then SKIP);Join$
    \end{tabular}\\
    \hline
\end{tabular}
\label{tab:join-csp}
\end{table}
\begin{table}[!ht]
\caption{Activity Final --- CSP Semantics}
\centering
\begin{tabular}{|c|}
	\hline
	\begin{tabular}{c}
	\includegraphics[scale=0.45]{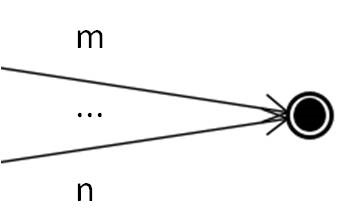}  
	\end{tabular} 
    \\
    \hline
    \begin{tabular}{c}
        $ActivityFinal = (ce.m \then SKIP \extchoice ce.n \then SKIP); clear \then SKIP$
    \end{tabular}\\
    \hline
\end{tabular}
\label{tab:activity-final-csp}
\end{table}

\begin{table}[!ht]
\caption{Merge --- CSP Semantics}
\centering
\begin{tabular}{|c|}
	\hline
	\begin{tabular}{c}
	\includegraphics[scale=0.45]{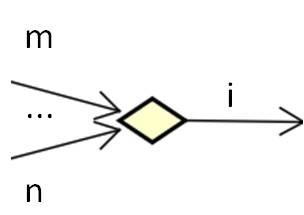}  
	\end{tabular} 
    \\
    \hline
    \begin{tabular}{c}
        $Merge = (ce.m \then SKIP \extchoice  ce.n \then SKIP); ce.i \then Merge$  
    \end{tabular}\\
    \hline
\end{tabular}
\label{tab:merge-csp}
\end{table}
\begin{table}[!ht]
\caption{Decision --- CSP Semantics}
\centering
\begin{tabular}{|c|}
	\hline
	\begin{tabular}{c}
	\includegraphics[scale=0.45]{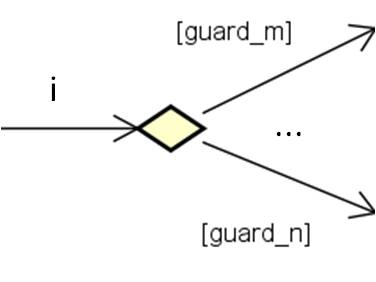}  
	\end{tabular} 
    \\
    \hline
    \begin{tabular}{ll}
        $Decision = ce.i \then $ & $(guard\_m$ \& $dc \then ce.m \then SKIP $\\
        & $ \extchoice ... $\\
        & $\extchoice guard\_n$ \& $dc \then ce.n \then SKIP) \backslash \{dc\}$  \\
        & $; Decision$ 
    \end{tabular}\\
    \hline
\end{tabular}
\label{tab:decision-csp}
\end{table}

Regarding object nodes, we provide semantics for parameter nodes, pins and object nodes (Tables \ref{tab:inParam-csp} and \ref{tab:outParam-csp}). Parameters and pins can be either input or output. An input parameter node has only outgoing edges while output parameter nodes just have incoming edges. When an activity is invoked, any values passed on parameters are wrapped in objects and placed on the activity parameter nodes at the start of the activity execution. In the corresponding CSP semantics, we use auxiliary memory process to store these values received in the $startActivity$ channel, which triggers the activity execution. These memory processes have $get\_[parameterName]$ and $set\_[parameterName]$ channels to read and update theses values. This process synchronizes on these channels with the process of the parameter node. Hence, input parameters simply read the value received by the activity and pass them to their outgoing edges in interleaving (Table~\ref{tab:inParam-csp}). Objects may flow into the output parameter nodes during the course of the execution of an activity. When the execution of the containing activity completes, the values contained in the output parameter nodes are passed out as outputs of the activity. This is mapped in CSP as external choice of the incoming edge events followed by the update of the memory process related to that parameter using the $set\_[parameterName]$ channel (Table~\ref{tab:outParam-csp}). 
Object nodes and pins are defined in a similar way using a memory process to store the value received in the incoming edge (using the $set$ channel). Then, we read the value (using the $get$ channel) out of the memory process and communicate it through the outgoing edge event of the node.
\begin{table}[!ht]
\caption{Input Parameter --- CSP Semantics}
\centering
\begin{tabular}{|c|}
	\hline
	\begin{tabular}{c}
	\includegraphics[scale=0.45]{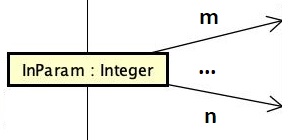}  
	\end{tabular} 
    \\
    \hline
    \begin{tabular}{ll}
        $InParam =$ & $update.(\#outEdges) \then get\_InParam?x \then $\\
        & $(oe.m.x \then SKIP \interleave ... \interleave oe.n.x \then SKIP)$\\
        $Mem\_InParam(x) = $ & $getInParam!x \then Mem\_InParam(x)$\\
        & $\extchoice setInParam?y \then Mem\_InParam(y)$
    \end{tabular}\\
    \hline
\end{tabular}
\label{tab:inParam-csp}
\end{table}
\begin{table}[!ht]
\caption{Output Parameter --- CSP Semantics}
\centering
\begin{tabular}{|c|}
	\hline
	\begin{tabular}{c}
	\includegraphics[scale=0.45]{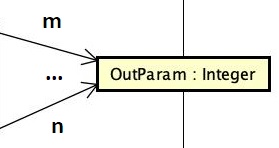}  
	\end{tabular} 
    \\
    \hline
    \begin{tabular}{l}
        $OutParam = oe.m?x \then set\_OutParam!x \then update.(-1) \then OutParam$\\
        \hspace{35pt}$\extchoice ... \extchoice oe.n?x \then set\_OutParam!x \then update.(-1) \then OutParam$\\
        $Mem\_OutParam(x) = getOutParam!x \then Mem\_OutParam(x)$\\
        \hspace{35pt}$\extchoice setOutParam?y \then Mem\_OutParam(y)$
    \end{tabular}\\
    \hline
\end{tabular}
\label{tab:outParam-csp}
\end{table}

Examples of CSP specifications generated from activity diagrams, including the ones shown in this paper, are provided in~\cite{Lima2019Supplementary}.

\section{Tool Support}
\label{sec:toolsupport}



A video illustrating installation, configuration and usage of the main framework features can be found in~\cite{Lima2019VerifierTool}.

\begin{figure}[h]
\centering
\includegraphics[scale=0.37]{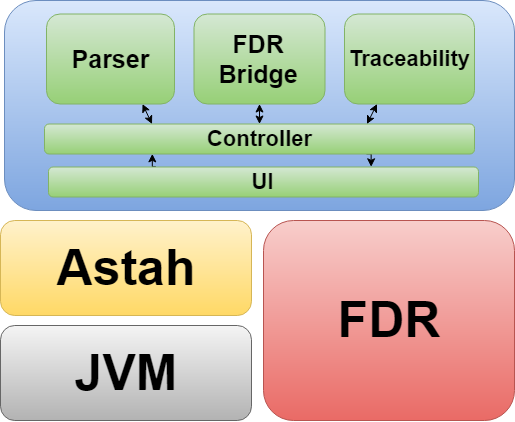}
\caption{Plug-in Architecture.}
\label{fig:architecture}
\end{figure}

Our framework has been implemented as a plug-in for the Astah modeling environment~\cite{Astah}. Figure~\ref{fig:architecture} shows how the plug-in and its dependencies are organized. We present the architecture in bottom-up order. Astah has several versions, which differentiate according to capabilities and languages supported. Our framework is built on top of its UML version. Such an environment allows extensions by the addition of plug-ins to add new features and runs on top of the JVM. The JVM (Java Virtual Machine), a virtual machine that allows a computer to run Java programs. It enables Java programs to be platform independent. Astah provides an API in the Java Programming Language to build its plugins. We use the Astah Java API to programmatically read and verify activity diagrams.
%
%
The plug-in depends on the FDR tool, which verifies CSP specifications. 

The developed plug-in is divided into modules, which are: UI, Controller, Parser, FDR Bridge and Traceability. The UI module is responsible for making the connection between the user and the controller through the plug-in menu. The Controller module is responsible for receiving information (commands and diagrams) from the UI module, managing the entire plug-in operation, and returning a response (messages and/or diagrams) to the UI module. The Parser module is responsible for receiving a diagram from the Controller module, translating it according to the semantics described in Section~\ref{sec:semantics}, and returning a CSP file to the Controller module. The FDR Bridge module is responsible for communicating with FDR, using the Java Reflection technique, which allows us to discover methods and attributes of a class at runtime, being able to load the FDR API dynamically. Finally, the Traceability module is responsible for receiving an event list (trace) of the Controller module, creating an activity diagram that shows the path traversed by the trace, and returning the diagram created to the Controller module.


\subsection{Activity diagram parser}

The process of translating the diagram is done automatically by the plug-in. The translation rules are encoded directly in the Java program we have built. This procedure may introduce inconsistencies as mistakes can be made during the coding tasks. However, to minimize this issue, the implementation process followed a Test-Driven Development approach~\cite{TDD} where first we define the test cases describing how the translation should be, and, next, we implement the parser code. Elements of the CSP specification for the input diagram are created in this order: processes that represent the nodes, definition of channels, definition of synchronization sets, main process, data types, Token Manager process and auxiliary memories processes.


During the translation of nodes we store all data types used by the nodes. Stored data is used in the subsequent steps of translation. This step traverses all nodes by creating a single process for each one and creating their synchronization set based on channels for incoming and outgoing edges. 
Such synchronization channels are responsible for making the connections between the processes. Each synchronization channel event represents an edge between two nodes. The main process translation step is responsible for starting the diagram, terminating the diagram and synchronizing sets of processes with the main process.

After translating all channels, we define the data types used by these channels according to the needs of the diagram. Next, we define the \textit{Token Manager} as described in Section~\ref{sec:semantics}. Finally, for nodes that require the storage of values, we define auxiliary memory processes.


\subsection{FDR integration}


Integration with FDR is very simple: 
the user has to inform the path to the FDR installation folder. 
This can be performed by accessing the 
plug-in UI menu 
\texttt{Tools -> Properties Plugin Configuration -> FDR Location}.

After informing the path, whenever a deadlock check or nondeterminism is required, the plug-in (FDR Bridge module) invokes FDR dynamically. Therefore, we do not need to include FDR as part of the plug-in.

\subsection{Deadlock and nondeterminism analysis}

\begin{figure}[h]
\centering
\includegraphics[scale=0.22]{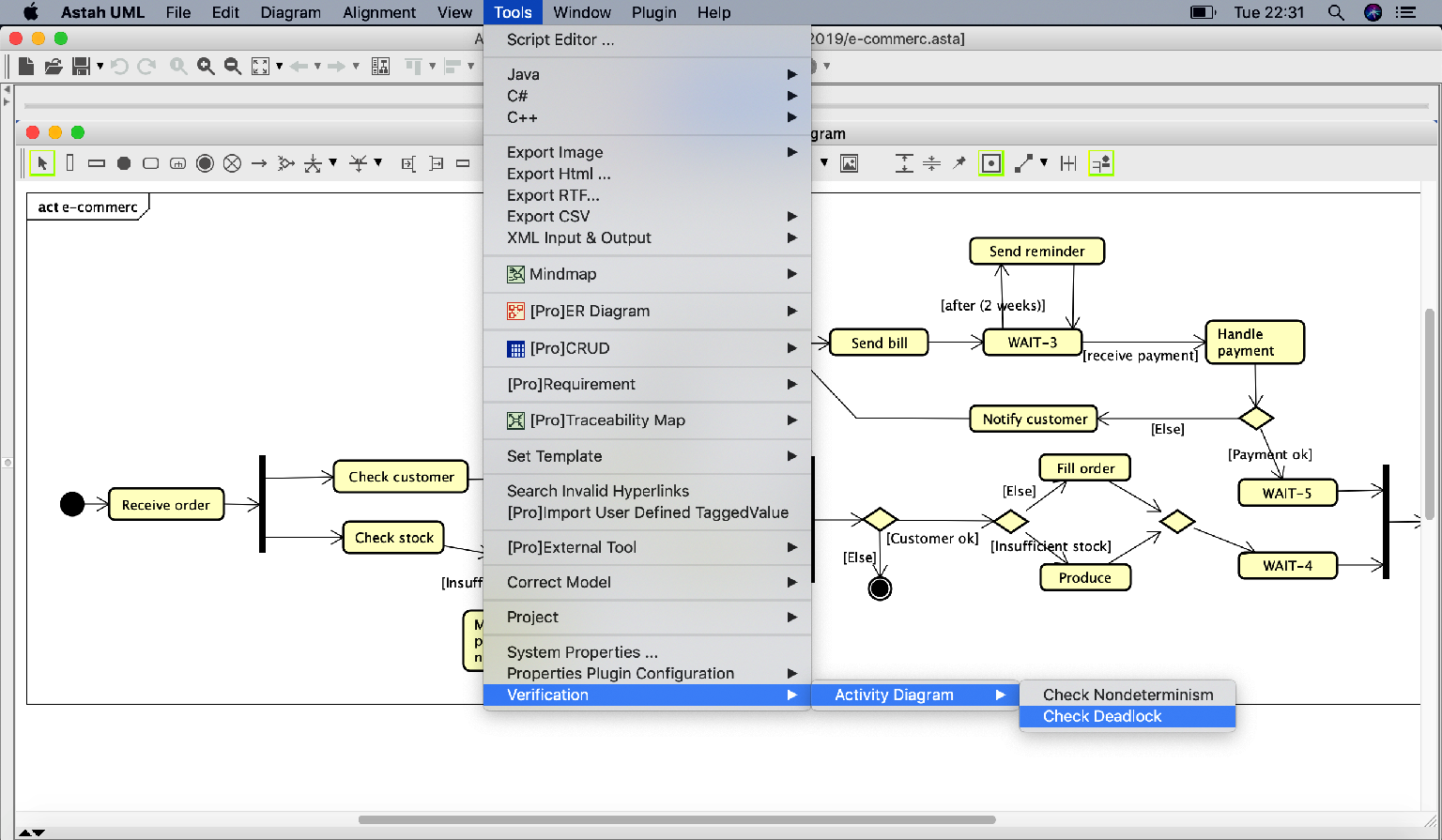}
\caption{Menu to choose the property to be verified.}
\label{fig:plugin-in-astah}
\end{figure}
\begin{figure}[h]
\centering
\includegraphics[scale=0.47]{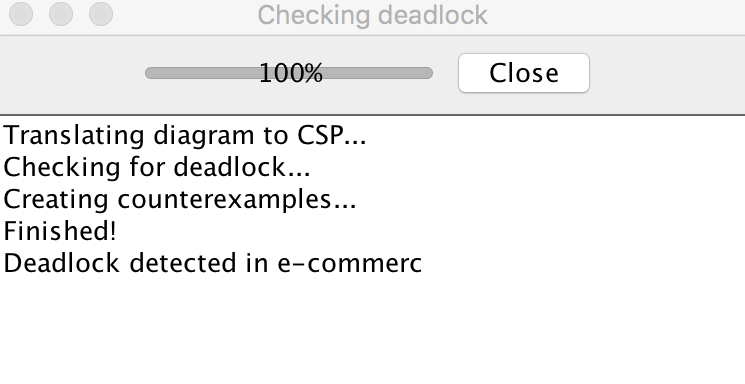}
\caption{Window showing the progress of the verification.}
\label{fig:verification-progress}
\end{figure}

In order to check if an activity diagram is deadlock free (or deterministic), the user only needs to click on the respective option displayed on the user interface. Figure~\ref{fig:plugin-in-astah} shows a menu that appears in the Astah tool after installing the plug-in. Such a menu allows the user to choose the type of verification to be performed in the current activity diagram. For instance, to perform deadlock freedom verification, one needs to access the plug-in UI menu \texttt{Tools -> Verification -> Activity Diagram -> Check Deadlock}. This action triggers the following tasks: the tool generates the corresponding CSP specification, loads it in FDR and invokes the deadlock-freedom assertion (deterministic assertion). Figure~\ref{fig:verification-progress} displays the progress window produced by the plugin for the verification of deadlock freedom, when the input diagram contains a deadlock behavior.
If a deadlock (nondeterministic behavior) is found by FDR, it returns a counterexample 
trace displaying the sequence of events that led to the deadlock (nondeterminism).



\subsection{Traceability}

\begin{figure}[h]
\centering
\includegraphics[scale=0.47]{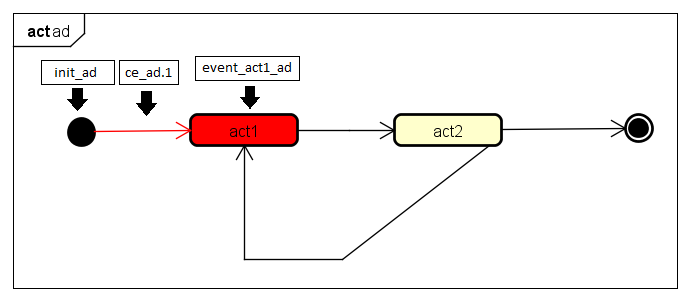}
\caption{Counterexample in Astah.}
\label{fig:counterexample}
\end{figure}

The trace returned by FDR when detecting a deadlock or nondeterminism is, as previously mentioned, an ordered list of events that shows us the path to where a deadlock or nondeterminism occurs. This ordered list is very important to identify where the problem occurred and also to create a diagrammatic counterexample that shows the issue in a view commonly known by the user. For example, in a deadlock check, given a trace returned
[\texttt{init\_ad}, \texttt{ce\_ad.1}, \texttt{event\_act1\_ad}], it is possible to identify the path traveled by FDR and where the deadlock occurred (in the event related to channel \texttt{event\_act1\_ad}). In other words, after reaching the event \texttt{event\_act1\_ad}, the process deadlocks. Figure~\ref{fig:counterexample} shows the traceability between the FDR counterexample and the diagram path that shows the deadlock. The nodes and flows painted red are the ones related to the events displayed in the trace returned by FDR. The diagram exhibited in Figure~\ref{fig:counterexample} is generated by the plugin, and put inside the current Astah project. Each verification counterexample produces a diagram inside Astah that highlights in red the path to the counterexample.

\section{Case Study}
\label{sec:casestudy}

This section shows the practical application of the proposed plugin to detect deadlock and nondeterminism in three different models in the domain of cloud computing systems.


Cloud computing is a modern approach to
build and remotely manage computing resources. 
A cloud, in the given context, refers to a complex, internet-based
infrastructure of hardware and software components~\cite{Maggiani09}.
It provides a variety of services such as: Software-as-a-Service (SaaS)
where software applications running on a cloud infrastructure are available as services to the consumers; Platform-as-a-Service (PaaS) in which development platforms and technologies are delivered as services to the users; and Infrastructure-as-a-Service (IaaS) where hardware elements like processing, network, storage and other fundamental computing resources are accessible for deploying and executing software.


Cloud computing is intimately tied to parallel and distributed computing. Thus, it is prone to common errors in these architectures. Among them, deadlock emerges as a crucial concern because it may be considerably hard to detect with the high number of components and relationships between them in a cloud computing environment. Another aspect refers to deterministic execution of tasks, which is usually a desirable property on systems. However, these systems are becoming incredibly complex and difficult to manage. Due to this fact,~\cite{Marinescu2013} proposes a question: should we migrate from a strictly deterministic view of such complex systems to a nondeterministic one? Therefore, we must have methods to assure the presence or absence of such properties. 

\begin{figure}[h]
\centering
\includegraphics[scale=0.41]{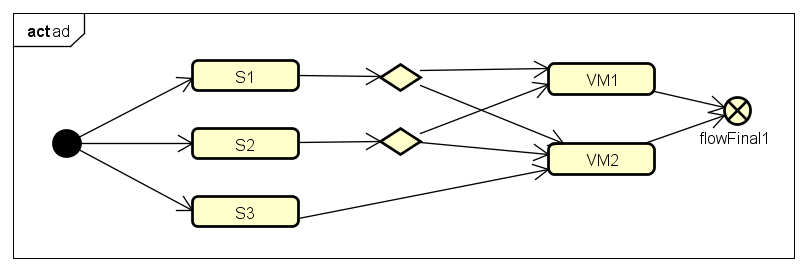}
\caption{Simple cloud computing network diagram.}
\label{fig:cloudcomputingnetwork}
\end{figure}

The first case is a simple cloud computing network model \cite{Cuong}, where S1, S2 and S3 are resources, and VM1 and VM2 are representations of machines in the cloud. The problem is related to allocation of resources to machines, where only one machine can use the resource at a time. Resources S1 and S2 can be allocated by both VM1 and VM2, and feature S3 can be allocated only by VM2. Figure \ref{fig:cloudcomputingnetwork} shows the diagram representing the network.

\begin{figure}[h]
\centering
\includegraphics[scale=0.42]{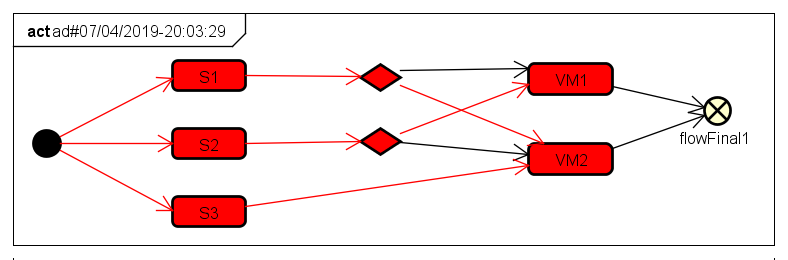}
\caption{Counterexample of cloud computing network deadlocked diagram.}
\label{fig:deadlockcloudcomputingnetwork}
\end{figure}

Using the plugin's deadlock check feature, we can find a deadlock occurrence on this network. The traceability feature constructs the diagram of Figure \ref{fig:deadlockcloudcomputingnetwork}, which shows where the deadlock occurred. The red edges and nodes represent the trace returned as counterexample by the FDR tool.

VM1 and VM2 nodes only execute when all their inputs are available. The cause of deadlock in this example is because both VMs need resources S1 and S2, resource S1 has been allocated to VM2 and resource S2 has been allocated to VM1 causing VMs to indefinitely wait for the other resource. Analyzing the counterexample diagram, we note that this problem could be solved by adding more resources to the network or by adding a central unit that manages the allocation of these resources, preventing VMs from disputing resources.

\begin{figure}[hbt]
\centering
\includegraphics[scale=0.42]{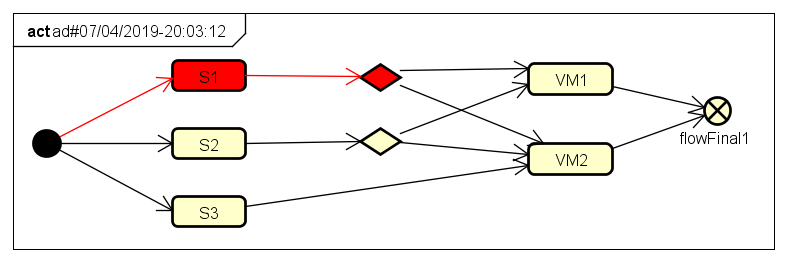}
\caption{Counterexample of cloud computing network nondeterministic diagram.}
\label{fig:determinismcloudcomputingnetwork}
\end{figure}

Figure \ref{fig:determinismcloudcomputingnetwork} shows the counterexample generated by the nondeterministic feature. The cause of nondeterminism in this example is that resource S1 can be allocated in either VMs, thus, there is not an explicit decision on which VM would have the resource. The red edges and nodes represent the trace returned by FDR, which shows the point where the nondeterminism occurs. What is interesting in this context is that this nondeterminism could be intentional. Hence, here we are assuring its presence. 

\begin{sidewaysfigure}
\centering
\includegraphics[scale=0.38]{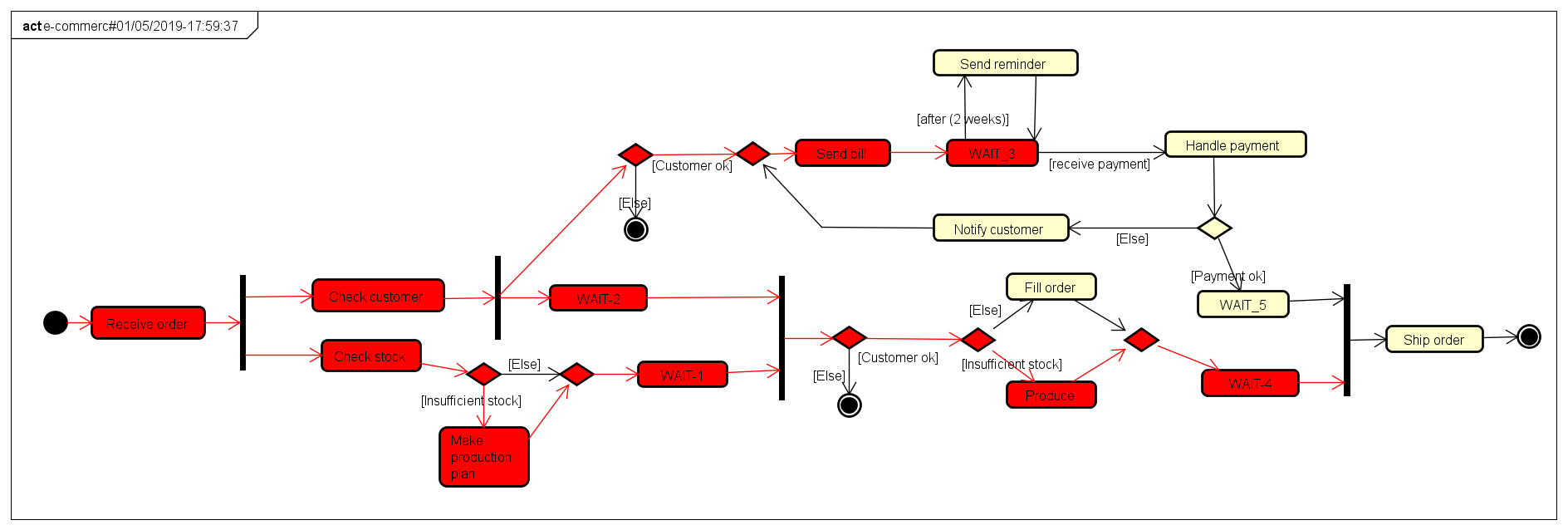}
\caption{Counterexample of e-commerce system diagram.}
\label{fig:deadlockecommerc}
\end{sidewaysfigure}

The second case is a model of an e-commerce system in the SaaS architecture \cite{Eshuis2006}, where sellers can create virtual stores to advertise their products. This functionality encompasses the customer order, production plan, production, payment and shipping the product. The problem here is the difficulty of manually ensuring that large models are free of deadlocks.
%
%
Figure \ref{fig:deadlockecommerc} shows the counterexample generated after the deadlock check. We can observe that between the \textit{WAIT-3} and \textit{Sender reminder} nodes there is a deadlock occurrence due to circular dependence between the nodes, making the \textit{WAIT-3} node wait by the output of the \textit{Sender reminder} node and vice versa.

\begin{figure}[htb]
\centering
\includegraphics[scale=0.50]{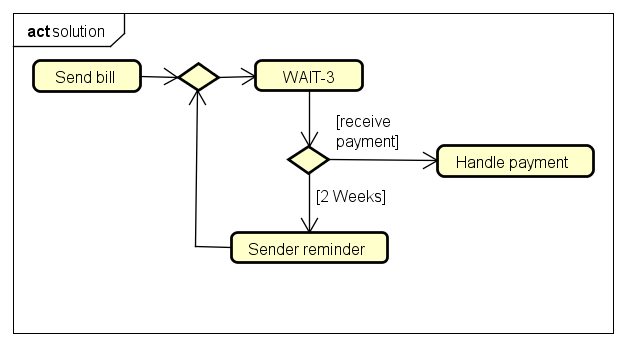}
\caption{Solution for deadlock of e-commerce system diagram.}
\label{fig:solutionecommerc}
\end{figure}

A possible solution to this case is to add a merge node in the incoming of \textit{WAIT-3}, which will receive the outputs of the \textit{Send bill} and \textit{Sender reminder}, and add a decision node to the outgoing of \textit{WAIT-3} to preserve the original loop semantics. Figure \ref{fig:solutionecommerc} illustrates the solution.

\begin{figure}[!htb]
\centering
\includegraphics[scale=0.50]{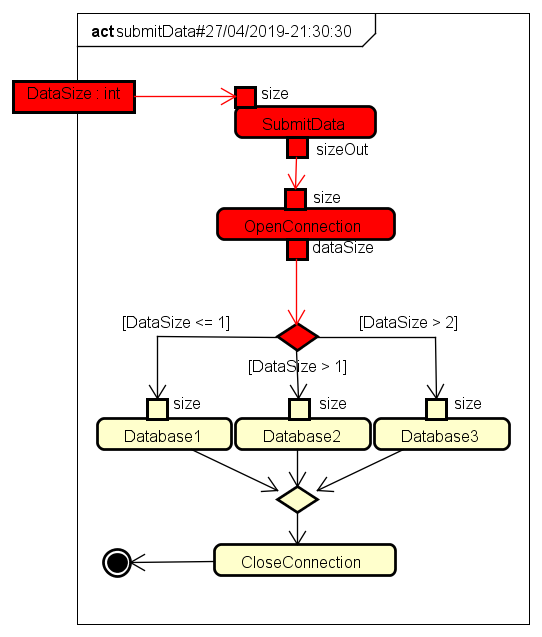}
\caption{Functionality of text file allocation on a cloud storage system.}
\label{fig:submitdata}
\end{figure}

Sometimes nondeterminism is a desired requirement for the system, so for the third case we show a feature of allocating text files in a cloud storage system in Figure \ref{fig:submitdata}. In this system, files less than or equal to 1MB are stored in Database 1, files greater than 1MB are stored in Database 2 and files having more than 2MB could be stored in databases 2 or 3. After checking for nondeterminism it is possible to show that the system actually has a nondeterministic choice to allocate text files in the last two databases. Red painted nodes and edges show the path to this nondeterministic decision on the diagram.




\subsection{Scalability Evaluation}

We have evaluated the scalability of our tool using five activities, including the ones shown in this paper. Table~\ref{tab:scalability} shows the models, their number of activity nodes, their number of activity edges, and the average times (in ms) of running ten deadlock analyses and ten nondeterminism analyses. Model \textbf{C1} corresponds to the activity shown in Figure~\ref{fig:cloudcomputingnetwork}, \textbf{C2} is the model for the file allocation example displayed in Figure~\ref{fig:submitdata}, \textbf{C3} is the motivation example of the paper (Figure~\ref{fig:deadlockedad}), \textbf{C4} describes an online hotel reservation system, and \textbf{C5} is the e-commerce system whose deadlock is illustrated in Figure~\ref{fig:deadlockecommerc}. We performed this evaluation in a machine with the following specification: 2.8 GHz Intel Core i5, 8 GB 1600 MHz DDR3, and 512 GB SSD Drive. 

\begin{table}[]
\caption{Scalability evaluation.}
\label{tab:scalability}
\begin{tabular}{ccccc}
\multicolumn{1}{l}{\textbf{Model}} & \multicolumn{1}{l}{\textbf{\#Nodes}} & \multicolumn{1}{l}{\textbf{\#Edges}} & \multicolumn{1}{l}{\textbf{Deadlock (ms)}} & \multicolumn{1}{l}{\textbf{Nondeterm. (ms)}} \\ \hline
\multicolumn{1}{c|}{C1}            & 9                                    & 12                                   & 642.6                                      & 672.9                                        \\
\multicolumn{1}{c|}{C2}            & 17                                   & 11                                   & 129.5                                      & 473.5                                        \\
\multicolumn{1}{c|}{C3}            & 10                                   & 11                                   & 1093.3                                     & 2336.4                                       \\
\multicolumn{1}{c|}{C4}            & 27                                   & 32                                   & 235.7                                      & 188.7                                        \\
\multicolumn{1}{c|}{C5}            & 32                                   & 37                                   & 33369                                      & 92117.7                                     
\end{tabular}
\end{table}

Table~\ref{tab:scalability} shows that the deadlock analysis is usually faster than the nondeterminism checking. The only exception is the C4 model, however, the difference is not significant.
Another aspect that can be perceived is that the number of nodes and edges not necessarily affects the complexity of the analysis. For instance, \textbf{C4} has more nodes and edges than \textbf{C1} and \textbf{C3} but its performance was better. The diagram \textbf{C4} does not present concurrent flows, however, \textbf{C1} and \textbf{C3} do have. We can conclude that not only the number of nodes and edges impacts the performance. The model concurrency level also influences the time to analyze the models. Therefore, nodes that may generate concurrent flows like forks (used in \textbf{C5}) and nodes with several outgoing edges (used in \textbf{C1} and \textbf{C3}) may significantly impact the performance of the analysis. Finally, we can see that even complex models like \textbf{C5} can be checked in approximately 30 seconds for deadlock and 90 seconds for nondeterminism. 

\section{Related Work}
\label{sec:relatedwork}

In this section, we cite works related to the verification of activity diagrams with the objective of analyzing properties such as refinement, deadlock, nondeterminism, and others.

The method proposed by George et al.~\cite{George2012} does the bad formulation check of activity diagrams, such as lack of borders, extra borders, decision node without guards, and others. The verification has two phases. First, the diagram is decomposed into its components. Then, each component is analyzed to prove whether it is well defined or not. The proposal is a simple algorithm to check whether the components are well defined or not. This work focuses on well-formedness of the activity diagram and not on properties like deadlock and nondeterminism. 

The method proposed by Eshuis~\cite{Eshuis2006} has two phases of translations. First, it converts the activity diagram to a hypergraph diagram and then converts the hypergraph diagram to a NuSMV specification that uses linear temporal logic (LTL) to specify its properties. The translation is based on the state machine and follows four steps: (1) insert a WAIT node for each edge joining a junction, (2) insert a WAIT node between a junction and a fork, (3) replace nodes and streams of objects by waiting nodes and control flows, (4) elimination of pseudo-nodes and definition of hyperedges. They propose to check for inconsistencies in activity diagrams during requirements and implementation phases. The method checks neither deadlock nor determinism. Furthermore, it does not have traceability mechanisms.

The method proposed by Elmansouri et al.~\cite{elmansouri_uml_2011} uses a meta-model to make a graphical transformation of an activity diagram (created in the ATOM tool) for a generalization of Chomsky's grammar for graphs. Then, it uses rules to translate Chomsky's diagram into a CSP specification. The proposal claims that the objective is to perform verification, but none are presented. In addition, the semantics of the constructors are not compositional, the number of constructors covered are not clear and does not seem to be as large as ours. 

In the method proposed by Alawneh et al.~\cite{Alawneh2006} a structure is developed to verify UML behavioral diagrams (state machine, activity and sequence diagrams). For each diagram, a formal semantic model is derived reflecting its characteristics and expresses its properties as formulas of temporal logic. The semantic model is called the Configuration Transition System (CTS), which are properties described in the temporal logic that are used as input to the NuSMV verification tool. Their proposal is to translate the diagram to CTS and check properties with the NuSMV tool. Although they do property checks, it lacks traceability back to the activity diagram, leaving the assessment and interpretation of problems to the user according to her experience on the formal notation.

The method proposed by Machida et al.~\cite{Machida11} describes a semi-automatic modeling structure called Candy, to quantify the availability of cloud services from the architectural and operational point of view. SysML~\cite{SysML1} activity diagrams are translated into components of the availability model and the components are assembled together to form the entire availability model in stochastic reward networks (SRNs). The generated availability model is evaluated from defined equations. This work
focuses on non-functional requirements regarding the availability of cloud computing systems, while we are interested in the functional aspect of these systems.

Baldan et al.~\cite{Baldan05} proposes a methodology for specification and verification of systems based on UML diagrams, which are interpreted in terms of graphs and graphical transformations. Once a system is modeled in this structure, a temporal-graphical logic can be used to express some of its relevant behavioral properties. Then, under certain constraints, such properties can be checked automatically. Although the verification is automatic, the transformation of the diagram is not. Moreover, evidences of traceability to the UML models are not presented.

Banti et al.~\cite{Banti2011} proposes a strategy to translate UML4SOA models into COWS terms to automatically check properties. Although the process does not require knowledge about the formal methods used, the method does not have an automatic form of traceability. In addition, the semantics of some terms, like the merge node, does not seem to be compositional.

The method proposed by Abdelhalim et al. \cite{Abdelhalim2010} translates a subset of fUML~\cite{fUML} to CSP in order to use FDR tool as a template checker. In this work, only the deadlock check is mentioned, whereas our work deals with nondeterminism as well as traceability back to the activity diagram.

The method proposed by Ouchani et al.~\cite{Ouchani2014} translates SysML activity diagrams into PRISM in order to check probabilistic properties using PCTL. The work does not cite verification of nondeterminism nor traceability of counterexamples to the SysML models.

Lima et al.~\cite{Lima2015} propose a formalization of several SysML diagrams, including activity diagrams, in terms of CML (COMPASS Modelling Language)~\cite{Woodcock2012a}, which is a formal language based on CSP and VDM for specifying systems of systems. This work inspired the definition of our formal semantics for activity diagrams. However, we had to adapt it to CSP and improve some definitions to achieve a better analysis performance and provide a suitable meaning to the models. Moreover, we increased the number of covered elements to augment expressiveness.


\section{Conclusions}
\label{sec:conclusions}


We have presented a framework for reasoning on properties of UML activity diagrams, more precisely, checking the presence of deadlock and nondeterminism. Such a framework uses a formal semantics defined in terms of the CSP process algebra that was based on previous works~\cite{Lima2013,Lima2015,LIMA2016}. We have adapted and incremented these initiatives in order to increase the expressiveness of the tool. It covers definitions for all types of nodes (action, control and object) and edges (control and object flows) following a compositional strategy, that is, each constructor has a semantic dissociated from others. This facilitates the mechanization of the strategy, the validation of the meaning of each element, and, the semantics can be easily extended. We take advantage of the wide range of CSP operators to support this compositionality.

Although CSP provides a rigorous and unambiguous language allied to a stable tool support (FDR), system and software engineers, generally, do not appreciate the manipulation of complex mathematical notations to design their models. Hence, one concern that has guided our framework is to avoid users having any contact with formal notations either to provide inputs or to read outputs from our tool. The only requirement is to inform the FDR installation directory. This is achieved by concealing all translations from UML models to CSP specifications added by the traceability mechanism from the output of FDR to UML models as well. We believe it is an attractive feature to UML practitioners in comparison to other works. In addition, allowing the validation of these properties in the same environment where the models are being built is another positive aspect of our strategy. 

Another distinctive feature of our framework is the nondeterminism checking. While deadlock is extensively discussed and covered in the literature, nondeterminism is a property that is usually set aside. However, if we consider the increasing complexity of systems and behaviors with too many decisions to be evaluated, this property becomes extremely relevant. That is why we have chosen to verify models related to cloud computing. Although our approach can be applied to other domains, models of cloud computing may deal with a large number of elements and relationships between them. Therefore, detecting the presence of properties may not be a trivial task. We have shown some examples with the detection of both deadlock and nondeterminism in activity diagrams related to this theme, however, we encourage the application of our framework in any scenario where the absence or presence of deadlock and nondeterminism must be ensured.

Although we have presented a scalability evaluation, the size of the models is not extremely large. Hence, we could not identify scenarios where our framework would not respond. We have an intuition that activities with a high level of concurrency or activities invoking several others may degrade the performance of the analysis. Therefore, we plan to stress these scenarios in future evaluations.

Nevertheless, there are some interesting topics for further research. Here we have only discussed applications related to deadlock and nondeterminism checking. However, other properties, especially related to the modeled domain, possibly are desired to be checked. Again, the problem is how to specify such properties. Approaches based on formal methods usually require the usage of some notation that can be related to the chosen formalism, based on some kind of logics like temporal logics (e.g., LTL, CTL~\cite{Baier2008} or PSL~\cite{PSL2}), or even a mathematical notation. However, they are not easy to specify by users with neither a mathematical nor a formal background. We envision the definition of properties in terms of another activity diagram enriched with some stereotypes and annotations. This would ease the adoption of the approach because we would not require the knowledge on another notation, only specific constructs to mark the activity diagram. For instance, the user would like to assure that a particular behavior happens inside a given activity diagram, or the opposite. The refinement theory of CSP and the refinement mechanisms of FDR give us the instruments to support the verification of such properties. 

In the current state, our strategy support basic types. As any bounded model checking mechanism, there are some cases where the state space may grow exponentially if the model deals with types with a large number of values. This can make the analysis unfeasible. To avoid that, we have created a mechanism where the user may restrict the interval of values to be considered in the analysis. In order to make this process less user-dependent, we plan to integrate our approach with solvers to automatically define the appropriate intervals.


The visual presentation of counterexamples is a relevant feature of the proposed approach. Nonetheless, there are opportunities to improve the way deadlock is presented to the user, so that the source for deadlock (and nondeterminism) becomes more clear and user-friendly. Improvements in the counterexample presentation will be addressed in future versions of the framework.

Our tool supports a considerable number of activity diagram constructors, including its most used ones. However, due to the compositionality aspect of our semantics, we hope to extend it with more constructors to provide as much expressiveness as possible to the users of our framework. 

A more theoretical line for future work is the usage of the precise semantics defined by fUML~\cite{fUML} to establish consistency between fUML and our CSP semantics, for the constructs covered by fUML. Exploring the relationship between fUML and CSP, in the context of the Unifying Theories of Programming~\cite{Hoare1998a}, which are used to give the a denotational semantics for CSP, is a promising way forward. In addition, we plan to formalize the translation rules using a notation that is programming language free.

Finally, we plan to develop more case studies in different domains to explore the semantic mapping and the reasoning strategy described here. Further studies to assess the scalability and usability of our approach are also in our plans of future research.

\section*{References}

\bibliography{bibliography.bib}

\end{document}